\documentstyle[12pt,epsfig]{article}             
\newcommand{\beq}{\begin{equation}}               
\newcommand{\eeq}{\end{equation}}                 
\newcommand{\bqry}{\begin{eqnarray}}              
\newcommand{\eqry}{\end{eqnarray}}                
\newcommand{\bqryn}{\begin{eqnarray*}}            
\newcommand{\eqryn}{\end{eqnarray*}}              
\newcommand{\preprint}[1]{\begin{table}[t]        
            \begin{flushright}                    
            \begin{large}{#1}\end{large}          
            \end{flushright}                      
            \end{table}}                          
\newcommand{\PD}[2]                               
    {\frac{\partial^{#2}}{\partial #1^{#2}}}      
\begin{document}
\preprint{\small hep-ph/9812395 \\
\small  LA-UR-98-5729}
\title{Generalized Schwinger Mass Formula}
\author{\\ L. Burakovsky,\thanks{E-mail: BURAKOV@T5.LANL.GOV} \
\ P.R. Page\thanks{E-mail: PRP@LANL.GOV} \
and \ T. Goldman\thanks{E-mail: GOLDMAN@T5.LANL.GOV}
\\  \\  Theoretical Division, MS B283 \\ Los Alamos National Laboratory \\ 
Los Alamos, NM 87545, USA \\}
\date{}
\maketitle
\begin{abstract}
We generalize Schwinger's original mass formula to the case of an additional 
isosinglet  mixing with the nonet mesons. 
We then make further generalization to either (i) an arbitrary number of 
additional isosinglets mixing with nonet mesons, or (ii) an arbitrary 
number of mesons  mixing with an additional isosinglet.
In the former case, we present an explicit relation, while in the latter 
case, we show by numerical example that the new mass formula is only weakly 
affected by the inclusion of additional mesons, and hence holds with good 
accuracy for each of the $3\times 3$ mass sub-matrices.  
\end{abstract}
\bigskip
\bigskip
{\it Key words:} Schwinger's formula, mass matrix, isosinglet-mesons mixing 

\noindent {\it PACS:} 12.39.Mk, 12.40.Yx, 12.90.+b, 14.40.-n
\bigskip
\newpage
\section{Introduction}
Schwinger's original mass formula \cite{Sch}  has attained celebrity 
due to its generality in a universe with only nine mesons, under the 
assumption that up, down and strange quarks may be treated the same.

If there were only up and down quarks which interact identically via the 
strong force, the $u\bar{u}$ and 
$d\bar{d}$ states of the hadron spectrum could be resolved into degenerate
isovector and isosinglet states\footnote{In the pseudoscalar
sector the axial anomaly and instantons can lift the degeneracy
\protect\cite{deriv}. In the scalar sector only instantons lift the
degeneracy, and in other sectors these effects are not significant
\protect\cite{deriv}.}. Although this
requires more than isospin invariance, we know it is consistent (for
mesons) with the details of QCD dynamics as there are no spin-isospin
correlations here to produce color magnetic spin interaction induced
splittings, at least for the lowest Fock components. 

The 
isovector masses can be regarded as physical since they undergo no further 
mixing. Therefore, in the Schwinger formula, the primitive (bare)
isosinglet masses can be equated with the (physical) isovector masses. 
Isosinglet states are invariant under rotation of up into down quarks and have
the flavour structure $n\bar{n}\equiv (u\bar{u}+d\bar{d})/\sqrt{2}.$
Enter the strange quark: Primitive $s\bar{s}$ states, the masses of which
are not {\it a priori} known, can mix with the primitive $n\bar{n}$ states to 
create physical states. Schwinger's formula pertains to the case where there 
is one $n\bar{n}$ and one $s\bar{s}$ primitive state, and where the strange 
quark behaves the same as the up and down quarks when the two primitive 
states couple to each other (SU(3) symmetry).

The formula has the benefit, except for the primitive $s\bar{s}$ mass, 
of referring only to experimentally measurable meson masses, 
and not intractable couplings. 

In this Letter we extend Schwinger's formula to include additional  
isosinglets and relax the SU(3) assumption of Schwinger.
This enables a more model independent analysis of the 
presence of isosinglets. In the case where the isosinglet is built mainly 
from gluons (a glueball), this extension is of central significance as some 
experimentally discovered isosinglet zero angular momentum states \cite{CBar}
are very widely accepted to exhibit mixing with a glueball 
\cite{review,weingarten}. 

Schwinger's formula 
\beq
\Big[ 3\tilde{M}(s\bar{s})-2M(s\bar{s})-M(n\bar{n})
\Big] \Big[ 3\tilde{M}(n\bar{n})-2M(s\bar{s})-M(n\bar{n})\Big]
+\;\!2\Big[ M(s\bar{s})-M(n\bar{n})\Big] ^2=0,
\eeq
relates the masses of the physical $(\tilde{M}(s\bar{s}),$ $\tilde{M}(n\bar{
n}))$ and primitive $(M(s\bar{s}),$ $M(n\bar{n}))$ mesons.
In this formula, and in other relations below, $M$'s 
stand either for the mass {\it or} mass squared.

The formula may be derived in the following way \cite{deriv}:
For a meson nonet, the isosinglet $q\bar{q}$ mass matrix modified by the 
inclusion of the mixing amplitudes $A,$ which are here assumed 
flavor-symmetric, for simplicity,
\beq
{\cal M}^{(2\times 2)}=\left( 
\begin{array}{cc}
M(s\bar{s})+A & \sqrt{2}A \\
\sqrt{2}A & M(n\bar{n})+2A
\end{array}
\right ) ,
\eeq 
is diagonalized by the masses of the physical mesons: ${\cal M}^{(2\times 2)}
\Longrightarrow {\rm diag}\left( \tilde{M}(s\bar{s}),\tilde{M}(n\bar{n})
\right) \!.$ Writing down the trace and determinant conditions for the 
matrix (2) and eliminating $A$ from them leads to the formula (1).
 
In Table I we compare 
the predictions of Schwinger's
 formula for the mass of the primitive $s\bar{s}$ 
state, based on the masses of the primitive 
$n\bar{n}$ state assumed to be the 
mass of the isovector, and the two physical states, with the predictions of 
the quark model motivated linear mass
relation $M(n\bar{n})+M(s\bar{s})=2M(s\bar{n}),$
where we use the mass of the isodoublet for $M(s\bar{n}).$ 

Schwinger's formula reflects the limitations on changes of masses from
primitives to physical states that can occur only by mixing. To the
extent that one may identify masses of primitives, by extracting mixing
angles or using model assumptions (as above), 
the formula agrees well with data in
most cases.

We interject a comment on the primitives. They are
not simply the result of suppressing quark loop contributions. For
example, in quenched lattice QCD, OZI rule violating decays of
$s{\bar s}$ states into two light quark-antiquark pairs may still
occur. Perhaps a useful analogy may be found in the issue of
CP-violating kaon decays - there are both mass mixing induced and
intrinsic CP-violating decays. Thus here it is insufficient simply to
extract the mixing angle to the primitives by determining from the
experimental decay rates which combinations of the physical states
would not violate the OZI rule. One must allow for the analog of the
intrinsic rule violating decays. 

Because of this, it continues to be much easier to {\it define} primitives on
the basis of some model assumption, such as that the light isoscalar
primitive is equal in mass to the light isovector state.  Further, the 
notion that the light-strange mixed quark primitive averages the masses
of the pure light and pure strange primitives receives adequate support
from a wide range of model calculations dating back to the earliest
days of the quark model through to the present refined potential and
string models.
The success of the Schwinger formula with these definitions of the 
primitives establishes the existence of primitives with these properties.
At this stage the glueball is ``integrated out'' in the formula.

\begin{figure}[t]
\begin{center}
{\footnotesize
\begin{tabular}{|c|c|c|c|c|} \hline
 $J^{PC}$ & States & I & II & III  \\ \hline
 $1^{--}$ & $\rho ,$ $K^\ast ,$ $\omega ,$ $\phi $ & 
$1013.0\pm 0.5$ & $1014.5\pm 0.4$ & $1017.8\pm 5.8$   \\ \hline
 $2^{++}$ & $a_2(1320),$ $K_2^\ast (1430),$ $f_2(1270),$ $f'_2(1525)$ &
$1541.7\pm 5.3$ & $1539.4\pm 5.5$ & $1539.9\pm 10.6$  \\ \hline
 $2^{-+}$ & $\pi _2(1670),$ $K_2(1770),$ $\eta _2(1645),$ $\eta _2(1870)$ &
$1870.7\pm 26.0$ & $1869.1\pm 25.3$ & $1869\pm 24$     \\ \hline
 $3^{--}$ & $\rho _3(1690),$ $K_3^\ast (1780),$ $\omega _3(1670),$ $\phi _3(
1850)$ & $1863.1\pm 9.2$ & $1862.3\pm 9.0$ & $1863.2\pm 16.1$   \\ \hline
\end{tabular}
}
\end{center}
Table I. Comparison of the values for $M(s\bar{s})$ given by Eq. (1) 
for linear and squared masses (in columns I and II, respectively), and 
the quark model motivated relation discussed in the text for linear masses 
(in column III). All the mass values are given in MeV. All the mass values 
are taken from ref. \protect\cite{pdg}.
Uncertainties represent the combination of experimental uncertainties and
the scale of electromagnetic splittings. 
\end{figure}

The agreement of Schwinger's formula with data 
is, however, problematical, as the mixing must pass
through glueballs determined by QCD dynamics. 
Here we shall examine such mass relations by explicitly including the
nearest glueball intermediate states to describe the mixing 
and complete the phenomenological picture, noting that  when the glueballs
are near to the mesons one should
 not be able to ``integrate out'' their effects.
We show that if the single intermediate
glueball state is only modestly more massive than mesons in the light
sector, then the $2 \times 2$ and $3 \times 3$ matrices give almost
identical results.  This is to be expected in the large glueball mass
limit, but we find it develops surprisingly early - just above 2 GeV.
We describe how the situation generalizes in the presence of multiple meson
nonets.

The validity of Schwinger's original formula (1) should be expected 
to break down in cases when additional isosinglets mix strongly with the 
primitive mesons. Such additional isosinglets may be glueballs or
hidden flavor heavy quark mesons with the quantum numbers of the 
nonet mesons. In such 
cases, one has to consider a larger mass matrix, $N\times N,$ $N>2,$ for 
which an analog of the original formula (1) may be expected to exist. 

In this paper we derive such an analog of Schwinger's original mass formula 
(1) in the case of any number of additional isosinglets (we need not specify 
their nature for our present purposes).

\section{Generalized Schwinger mass formula}

Let us first note that a field theory in general, and QCD in particular, does
not have a  mass matrix corresponding to composite states. 
Indeed, the diagonalization of a mass 
matrix corresponds to a linear transformation of fields that diagonalizes the 
Lagrangian, which can only be done for Lagrangians which are
 quadratic in fields, 
e.g., the Klein-Gordon Lagrangian, with quadratic interactions. A mass matrix 
appears therefore only in {\it effective} field theories of this type. 
However, 
precisely this type of an effective meson theory may be expected to exist for 
QCD. In fact, the emergence of such a theory was shown in \cite{PRC} through 
the bosonization of QCD. Hence, we assume the existence of an effective 
theory that describes the mixing of nonet mesons with additional isosinglets.

We start with the case of one additional isosinglet. Assume, for simplicity, 
that an additional isosinglet (glueball or hidden flavor heavy quark meson), 
$i,$ couples to a pair of primitive (nonet) mesons. Then, we have the 
following $3\times 3$ hermitian mass matrix where, in addition to the quark 
mixing amplitudes, we have the amplitudes of the $i-q\bar{q}$ mixing
(the couplings) which we denote by $B:$  
\beq
{\cal M}^{(3\times 3)}=\left(
\begin{array}{ccc}
M(i) & B & Br \\
B^\ast  & M(s\bar{s})+A & Ar \\
B^\ast r^\ast  & Ar^\ast  & M(n\bar{n})+A|r|^2
\end{array}
\right) ,
\eeq 
where $M(i)$ is the mass of the primitive $i.$ From now on, in contrast to 
Eq. (2), we allow for possible violation of exact flavor SU(3) symmetry in 
terms of arbitrary $r.$ The SU(3) limit will then correspond to $r=\sqrt{2}.$ 
The form of this mass matrix is motivated in 
ref. \cite{Fuchs}. The mass matrix (3) was also discussed in \cite{Turnau}, 
and its SU(3) symmetric version in \cite{matrix}.

Note that in Eq. (3), as well as in Eq. (2), $A$ must be real. With $B$ and 
$r$ both real the matrix (3) is the most general parametrization
of $3\times 3$ (real) symmetric 
matrix, since it contains 6 independent parameters. 

The mass matrix (3) is diagonalized by the masses of the three physical 
states,
${\cal M}^{(3\times 3)}\Longrightarrow {\rm diag}\left( \tilde{M}(i),\tilde{
M}(s\bar{s}),\tilde{M}(n\bar{n})\right) ,$ which are determined from the three
eigenvalue equations (which follow from ${\rm Det}\left( {\cal M}^{(3\times 
3)}-\lambda I\right) =0).$ Eliminating $A$ and $B$ 
from the eigenvalue equations leads, upon some algebra, to the formula

$$\Big[ (1+|r|^2)\;\!\tilde{M}(i)-|r|^2\;\!M(s\bar{s})-M(n\bar{n})\Big] 
\Big[ (1+|r|^2)\;\!\tilde{M}(s\bar{s})-|r|^2\;\!M(s\bar{s})-M(n\bar{n})
\Big] $$ $$\times \Big[ (1+|r|^2)\;\!\tilde{M}(n\bar{n})-|r|^2\;\!M(s\bar{s})
-M(n\bar{n})\Big] $$
\beq
+\;\!\Big[ (1+|r|^2)\;\!M(i)-|r|^2\;\!M(s\bar{s})-M(n\bar{n})\Big] \;|r|^2\;\!
\Big[ M(s\bar{s})-M(n\bar{n})\Big] ^2=0,
\eeq
which is the generalized Schwinger mass formula for the case of a meson nonet 
with an additional isosinglet. We note  that the new formula does not depend 
on either the values of $A$ or $B,$ just as the original formula does not 
depend on the value of $A.$ 

\begin{figure}[t]
\begin{center}
\vspace{-1.5cm}
\epsfig{file=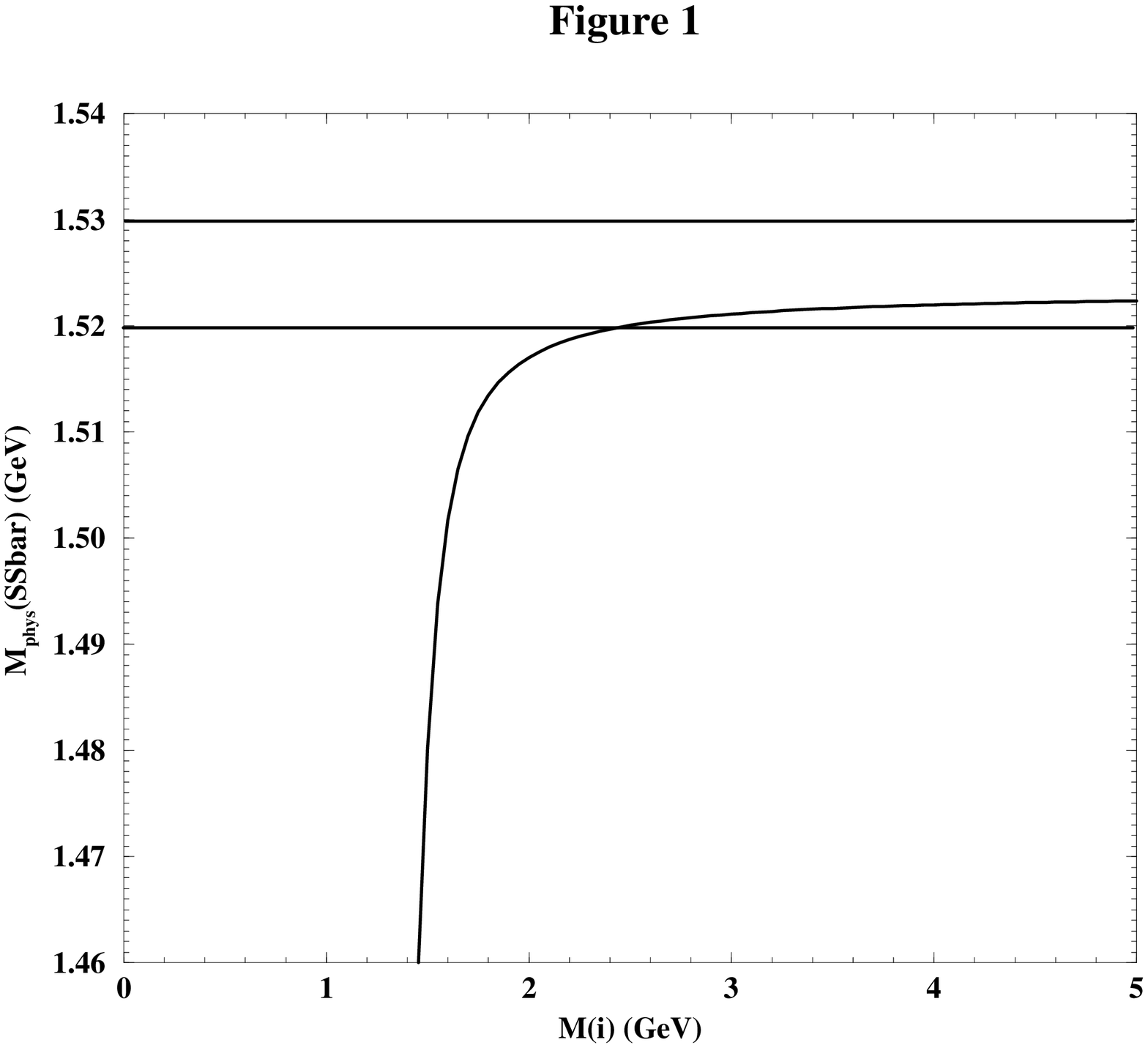,width=15cm,angle=-90}
\end{center}
Fig. 1. Comparison of the mass of the physical $s\bar{s}$ state 
calculated from Eqs. (4),(5) as a function of the bare mass of an additional
isosinglet, $M(i),$ with the range allowed by experimental data of ref. 
\cite{pdg} (horizontal band), for the $J^{PC}=3^{--}$ multiplet.\hfil\break 
\end{figure}

Under the conditions that either
 
(i) $(1+|r|^2)\;\!M(i)\gg |r|^2\;\!M(s\bar{s})+M(n\bar{n})$ (see Eq. (4)), or

(ii) $|B|\ll A,$ \\
so that $\tilde{M}(i)\approx M(i),$ the generalized Schwinger formula (4) 
(with $r=\sqrt{2})$ reduces to the original one, Eq. (1). 

In the case (iii) $A=0,$ \\
i.e., no direct coupling between mesons, one can combine the new formula (4) 
with one of the eigenvalue equations (the trace condition for the matrix (3))
\beq
\tilde{M}(i)+\tilde{M}(s\bar{s})+\tilde{M}(n\bar{n})=M(i)+M(s\bar{s})+
M(n\bar{n})
\eeq
in order to determine $two$ unknown masses. Under the assumption that (iii) 
is 
the realistic case for meson spectroscopy, i.e., that the observed meson mass 
spectra result from coupling of the primitive $q\bar{q}$ states through an 
additional isosinglet, one can calculate the masses of two physical states, 
e.g., $\tilde{M}(i),$ $\tilde{M}(s\bar{s}),$ as functions of $M(i),$ using the
known $M(n\bar{n}),$ $M(s\bar{s}).$ In Figs. 1,2 we compare $\tilde{M}(s\bar{
s})$ calculated in this way with the range allowed for $\tilde{M}(s\bar{s})$
by experimental data \cite{pdg}, for the $J^{PC}=3^{--}$ 
multiplet\footnote{The other meson nonets can be analyzed in a similar way. 
The case of the $J^{PC}=0^{++}$ has been studied in detail in ref. \cite{BP}.}
(here we use Eqs. (4),(5) for linear masses). One sees that the value 
calculated lies within the range allowed by data for a rather large bare
mass of an additional isosinglet, $\stackrel{>}{\sim }2$ GeV.

\begin{figure}[t]
\begin{center}
\vspace{-1.5cm}
\epsfig{file=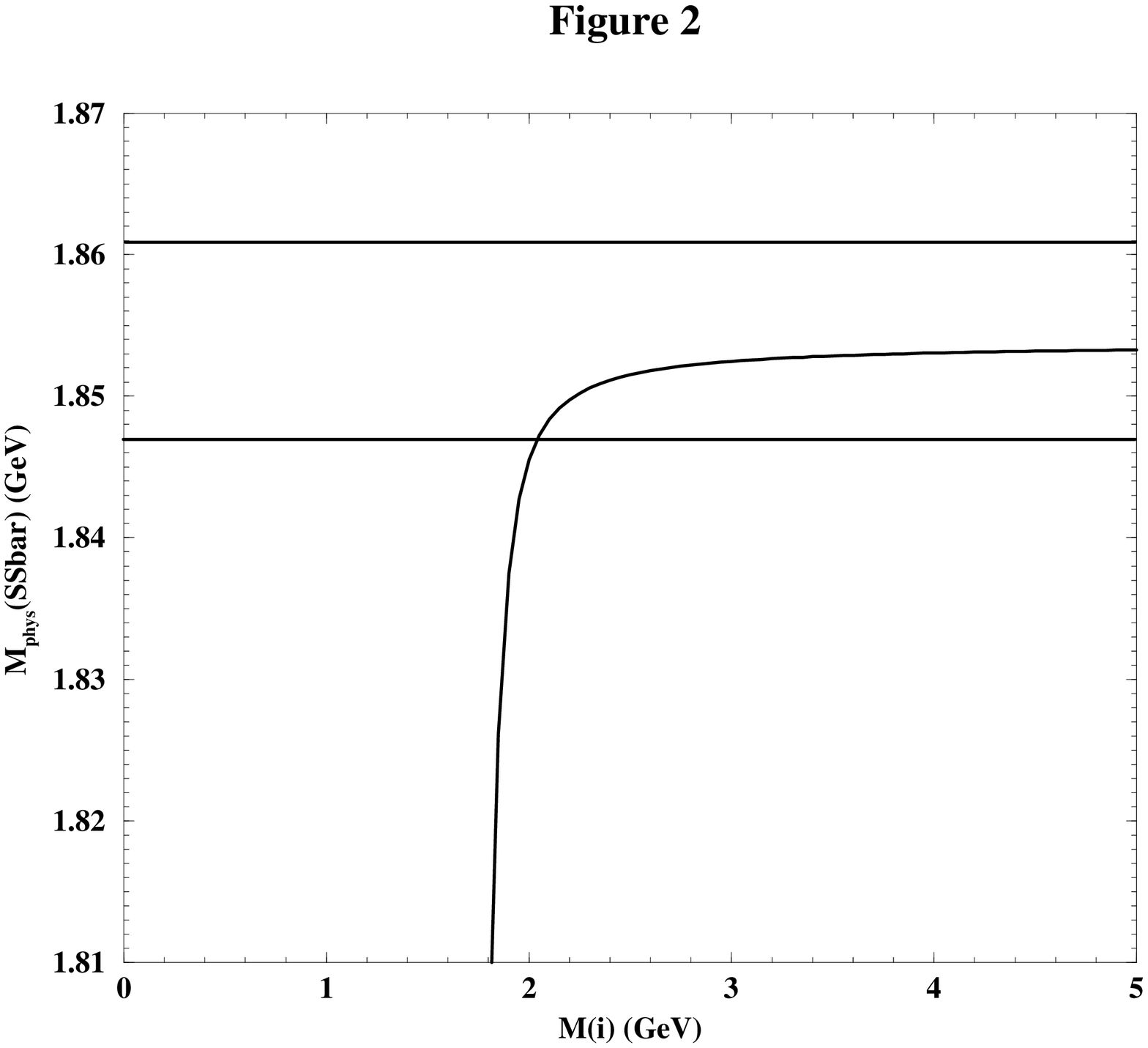,width=15cm,angle=-90}
\end{center}
\centerline{Fig. 2. The same as Fig. 1 for the $J^{PC}=3^{--}$ 
multiplet.\hfil\break }
\end{figure}

Note that $all$ isosinglet masses above 2.5 and 2 GeV, respectively, are 
consistent with $\tilde{M}(s\bar{s})$ from experiment, so that the mass of an
additional isosinglet cannot be uniquely determined from current experimental
data. Thus, the procedure described above (only) fixes the lower bound on the
mass of an additional state $(i).$

\section{Further generalization of the new Schwinger mass formula}
The most general case for a mass formula of the type (4) is evidently the
case of the mixing of arbitrary numbers of both mesons and additional 
isosinglets with common $J^{PC}$. This most general case can be decomposed 
into the following two cases: (i) an arbitrary number of additional 
isosinglets (glueballs and hidden flavor heavy quark mesons) mixing with 
one pair of (nonet) mesons built from $u,d,s$ quarks, 
and (ii) an arbitrary number of pairs of excited 
mesons (with common $J^{PC})$ built from $u,d,s$ quarks 
mixing with one isosinglet (glueball or hidden 
flavor heavy quark meson). We now examine these two cases in more detail.

\subsection{The case of an arbitrary number of additional isosinglets}
In this case the new Schwinger mass formula has an explicit further 
generalization, as follows. 

Consider the mixing of $(N-2)$ extra isosinglets with a pair of nonet mesons 
and introduce, as previously, the corresponding bare masses $M(i_1),\ldots ,M(
i_{N-2}),$ couplings $B_1,\ldots ,B_{N-2},$ quark mixing amplitudes $A$ and 
flavor SU(3) breaking parameter $r.$ The $N\times N,\;N\geq 3$ hermitian mass 
matrix is then diagonalized by the masses of the $N$ physical states:
\beq
\left( \begin{array}{ccccccccc}
  M(i_1) &   0   & \cdot & \cdot & \cdot &   0   &   B_1   &   B_1r   \\
   0  &  M(i_2)  & \cdot & \cdot & \cdot &   0   &   B_2   &   B_2r   \\
\cdot & \cdot & \cdot &      &       &  \cdot   &  \cdot  &  \cdot   \\
\cdot & \cdot &       & \cdot &      &  \cdot   &  \cdot  &  \cdot   \\
\cdot & \cdot &       &       & \cdot & \cdot   &  \cdot  &  \cdot   \\
   0  &  0  & \cdot & \cdot & \cdot &  M(i_{N-2}) & B_{N-2} & B_{N-2}r \\
  B_1^\ast  & B_2^\ast  & \cdot & \cdot & \cdot & B_{N-2}^\ast  &
  M(s\bar{s})+A  & Ar   \\
  B_1^\ast r^\ast  & B_2^\ast r^\ast  & \cdot & \cdot & \cdot & 
B_{N-2}^\ast r^\ast  & Ar^\ast  & M(n\bar{n})+A|r|^2
\end{array}
\right) 
\eeq
\beq
\Longrightarrow {\rm diag}\;\Big( \tilde{M}(i_1),\tilde{M}(i_2),\ldots ,
\tilde{M}(i_{N-2}),\tilde{M}(\bar{s}s),\tilde{M}(\bar{n}n)\Big) .
\eeq
Here we assumed 
that it is possible to choose a basis where the primitive 
additional isosinglet states are already diagonalized. For example,
if the three isosinglets are a glueball, $c\bar{c}$ and $b\bar{b}$, such
a basis can to a good approximation be chosen due to the small mixing
between the glueball and hidden flavour heavy quark mesons. We also 
assumed that the $n\bar{n}$ meson to $s\bar{s}$ meson coupling ratio of 
all isosinglets is a universal constant $r$. In the limit of SU(3) symmetry
this assumption becomes exact, and $r=\sqrt{2}$.

Since the initial $N\times N$ mass matrix (6) can be rewritten in the 
following effective $3\times 3$ form:
\beq
\left( \begin{array}{ccc}
    {\bf M}(i)    & {\bf B} & {\bf B}r   \\
  {\bf B}^\dagger  & M(s\bar{s})+A & Ar   \\
 {\bf B}^\dagger r^\ast  & Ar^\ast  & M(n\bar{n})+A|r|^2
\end{array} 
\right) ,
\eeq
where
\beq
{\bf M}(i)\equiv \left( \begin{array}{cccccc}
M(i_1) & 0 & \cdot & \cdot & \cdot & 0   \\
0 & M(i_2) & \cdot & \cdot & \cdot & 0   \\
\cdot & \cdot & \cdot &    &    & \cdot  \\
\cdot & \cdot &    & \cdot &    & \cdot  \\
\cdot & \cdot &    &    & \cdot & \cdot  \\
0 & 0 & \cdot & \cdot & \cdot & M(i_{N-2})
\end{array}
\right) ,\;\;\;{\bf B}\equiv \left( \begin{array}{c}
 B_1  \\
 B_2  \\
\cdot \\
\cdot \\
\cdot \\
B_{N-2}
\end{array}
\right) ,
\eeq
it is evident that the general $N\times N$ case effectively reduces to the 
$3\times 3$ case considered above. Applying the methods of linear algebra,
one can obtain upon some calculation the following most general Schwinger type
mass relation for the $N\times N$ case:\footnote{One of us (L.B.) wishes to
thank C.D. Levermore for a very useful discussion on this point.}
$$\prod _{k=1}^{N-2} \Big( (1+|r|^2)\tilde{M}(i_k)-|r|^2M(\bar{s}s)-
M(\bar{n}n)
\Big) \Big( (1+|r|^2)\tilde{M}(\bar{s}s)-|r|^2M(\bar{s}s)-M(\bar{n}n)\Big) $$ 
$$\times \Big( (1+|r|^2)\tilde{M}(\bar{n}n)-|r|^2M(\bar{s}s)-M(\bar{n}n)
\Big) $$
\beq
+\;\prod _{k=1}^{N-2} \Big( (1+|r|^2)M(i_k)-|r|^2M(\bar{s}s)-M(\bar{n}n)\Big) 
\;|r|^2\;\Big( M(\bar{s}s)-M(\bar{n}n)\Big) ^2=0.
\eeq
In the case of only one additional isosinglet, Eq. (9) reduces to the 
generalized Schwinger mass formula (4). 

\section{The case of an arbitrary number of mesons}
As discussed above, the generalized Schwinger mass formula (10) reduces 
to the original formula (1) in the limits where the primitive mass of an 
additional isosinglet goes to infinity, or the isosinglet-meson coupling 
goes to zero. The main use of the new Schwinger formula is hence where an 
additional isosinglet is in the vicinity of the mesons, where the
conventional Schwinger formula is expected to be inapplicable. 
However, as the two 
examples of the $J^{PC}=2^{++}$ and $3^{--}$ multiplets considered above show,
the mass of such an additional isosinglet must generally be expected to be 
large (in agreement with, e.g., lattice QCD calculations of glueballs
\cite{LQCD}). Hence the mesons in the new Schwinger formula would need to be 
radially or orbitally excited mesons, or hybrid mesons (in all nonets except 
the scalar nonet \cite{BP}). It now becomes interesting to consider whether 
the excited states could influence the mixing of an additional isosinglet with
the ground states. It should be decided how many pairs of excited states
to incorporate. 
There are arguments that the effective number of hadronic states of any given 
type is limited \cite{BBG2}. We shall discuss a 7-state case which is 
sufficiently complicated to simulate the general case. The $7\times 7$ matrix 
has also been discussed in a different context in ref. \cite{FOD}.
%

Here we investigate a set of three pairs of mesons and test for the effect 
by choosing an additional isosinglet at various masses. We choose ground 
states at $M(n\bar{n})=1.2$ GeV, $M(s\bar{s})=1.5$ GeV, first excited
states at $M^{(1)}(n\bar{n})=1.7$ GeV, $M^{(1)}(s\bar{s})=2$ GeV, and second 
excited states at $M^{(2)}(n\bar{n})=2.1$ GeV, $M^{(2)}(s\bar{s})=
2.4$ GeV. The mass splittings chosen are typical of constituent quark models
\cite{review}. The additional isosinglet mass cases we consider are (i)-(vi). 
Each of the following two cases,

(i) $M(i)=1.35$ GeV,

(ii) $M(i)=1.6$ GeV \\
simulates the scalar meson case for which an additional isosinglet (the scalar
glueball) is in the vicinity of the ground state mesons \cite{BP}. Each of 
the following three cases,

(iii) $M(i)=1.85$ GeV,

(iv) $M(i)=2.05$ GeV,

(v) $M(i)=2.25$ GeV \\
simulates the tensor meson case for which the mass splitting between the 
tensor glueball and the $n\bar{n}$ ground state meson is \cite{LQCD} 
$\sim 0.9$ GeV. Finally, the case

(vi) $M(i)=2.5$ GeV \\
simulates the $2^{-+}$ and $3^{--}$ cases for which the mass splitting between
the corresponding glueball and $n\bar{n}$ ground state meson is \cite{LQCD} 
$\sim 1.3$ GeV.

For each of the above six choices of the primitive mass of an additional 
isosinglet, we first consider the full $7\times 7$ mass matrix for the mixing 
of the isosinglet with ground states and both pairs of excited states which
generalizes the $3\times 3$ matrix (5). The couplings of an additional 
isosinglet to mesons are taken to be the same for ground states and both 
excited states, for simplicity:\footnote{This value represents the realistic 
value of the glueball-meson coupling, as extracted from data by the 
phenomenological analysis in ref. \cite{BBG}.} $B=0.4$ GeV$^2,$ and no direct 
meson-meson coupling\footnote{It can be shown that there are analytical 
Schwinger-type formulae in this case \protect\cite{bur99}.}: $A=0.$ 
We then consider separately the mixing
of an additional isosinglet with ground states and first and second  
excited states, each of which is described by the corresponding $3\times 3$ 
mass matrix with the same $B=0.4$ GeV$^2$ and $A=0.$ In each case, we 
calculate the masses of all the physical states, using the appropriate 
quadratic mass relations. Our results in the full $7\times 7$ case and the 
separate $3\times 3$ cases are compared in Table II.

As seen in Table II, the largest discrepancy between the results of the
$7\times 7$ and $3\times 3$ cases occurs for the $n\bar{n}$ ground state in 
the case (i) and constitutes $\sim 8$\%. In every other case, the 
corresponding discrepancy does not exceed $\sim 2$\% (which corresponds to a 
mass uncertainty of $<50$ MeV), which means that the generalized Schwinger 
mass formula in every $3\times 3$ sub-basis should be very accurate. We note 
that in the actual case of the scalar nonet with smaller $B=0.3$ GeV$^2$ 
\cite{BP}, in the case (i) one would obtain $\tilde{M}(n\bar{n})=1032$ MeV, 
$\tilde{M}(s\bar{s})=1535$ MeV for the $7\times 7$ case, and $\tilde{M}(
n\bar{n})=1068$ MeV, $\tilde{M}(s\bar{s})=1565$ MeV for the $3\times 3$ case, 
so that the discrepancy between the results for the $\tilde{M}(n\bar{n})$ 
reduces to only $\sim 3.5$\%.

\begin{figure}
\begin{center}
\vspace{-1.5cm} 
\epsfig{file=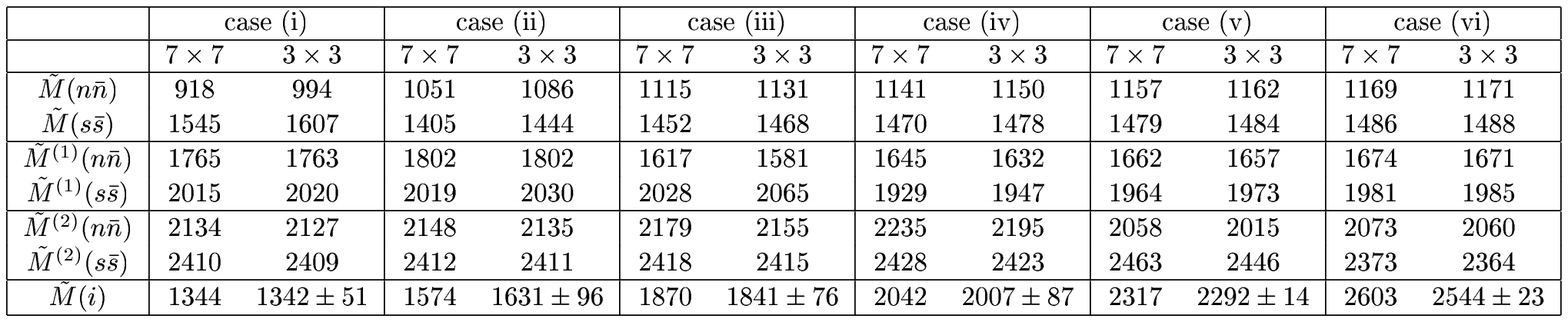,width=27cm,angle=90}
\end{center}
Table II. Comparison of the masses of the physical states in the full
$7\times 7$ and the separate $3\times 3$ cases, for each of the six choices 
of the primitive mass of an additional isosinglet. Only the average value of 
the physical isosinglet mass from the $3\times 3$ matrices is indicated. All 
masses are given in MeV.\hfil\break
\end{figure}

We have also done a similar comparison between the $7\times 7$ and $3\times 3$
cases for the same six choices of the primitive mass of an additional 
isosinglet and the same bare mesons masses, but with nonzero direct 
meson-meson couplings, $A,$ in each of the $2\times 2$ mesons
sub-bases, in addition to nonzero $B.$ For $A=B=0.4$ GeV$^2,$ the largest
discrepancy between the corresponding results again occurs for the $n\bar{n}$
ground state in the case (i) and constitutes $\sim 7$\%. In every other case,
the corresponding discrepancy does not exceed $\sim 5$\% (which corresponds to
a mass uncertainty of $<80$ MeV). 

Note that in the above analyses, for ground states and both pairs of excited 
states we formally used a common value of $B.$ 
However, the couplings of the excited states (to the glueball) may be smaller.
This would give rise to even smaller errors than quoted above.

On the basis of both possibilities for zero and nonzero $A$ analyzed, we 
conclude that the generalized Schwinger formula is only weakly affected 
by the inclusion of the excited states, and must therefore hold with
good accuracy for almost every $3\times 3$ isosinglet-mesons mass sub-matrix.

\section{Conclusions} 

Generalized Schwinger mass formulas have been derived for an arbitrary number
of isosinglets coupled to a meson nonet. The formulae have significant
utility as they mainly refer to experimentally measurable masses, and not 
unknown couplings; they can therefore be used to predict unknown masses.
For real couplings, the generalized Schwinger mass formula with the inclusion
of a single isosinglet describes the most general parametrization of
a symmetric $3\times 3$ matrix. In other cases, the matrices
are not the most general hermitian matrices, albeit physically relevant ones.

When multiple meson excitations were included in addition to a single
isosinglet, we considered the most general matrices. It has been shown for 
some numerical examples that to a good approximation it is sufficient to 
restrict to the relevant $3\times 3$ isosinglet-mesons subspace.


\end{document}